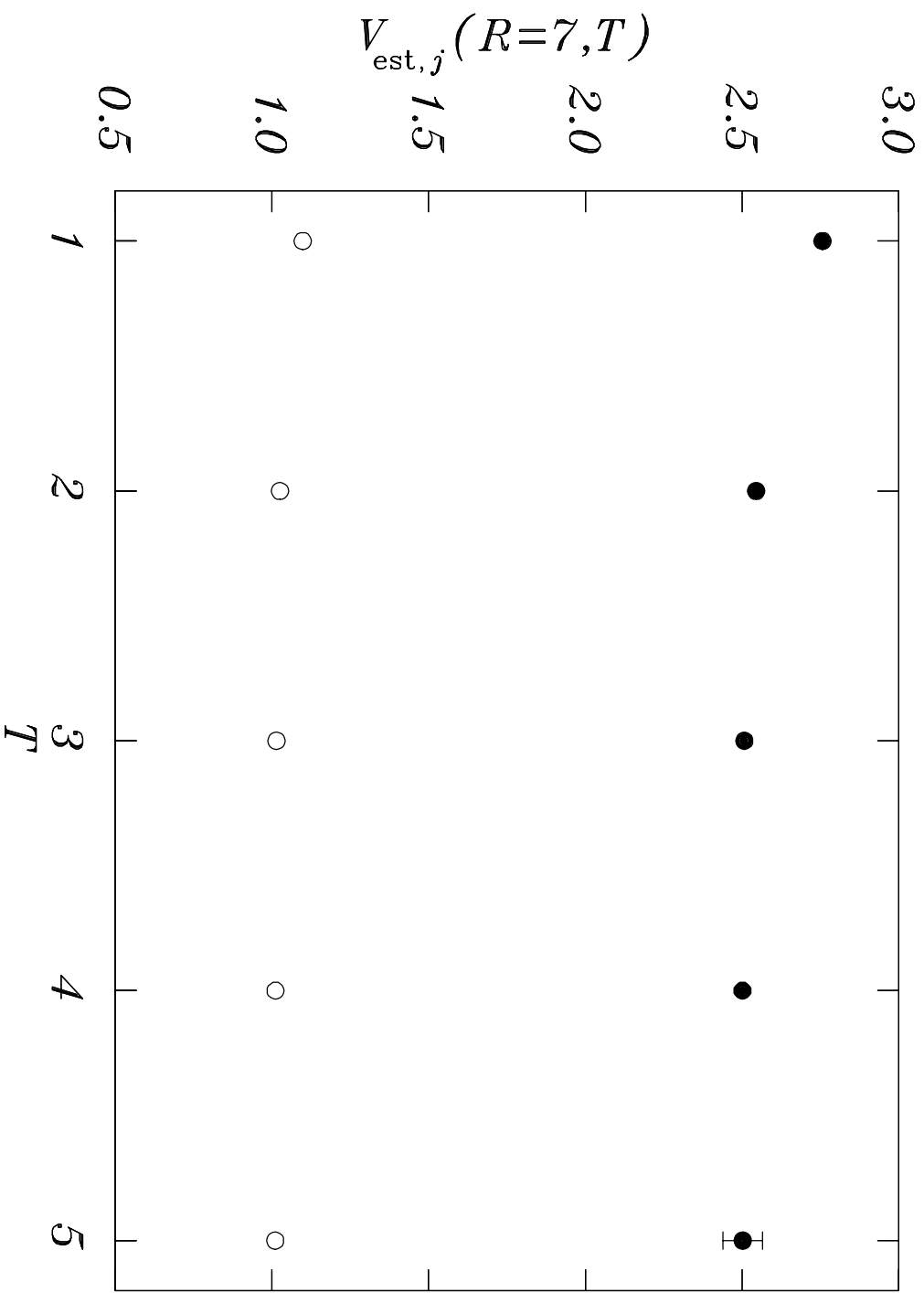

FIGURE 1

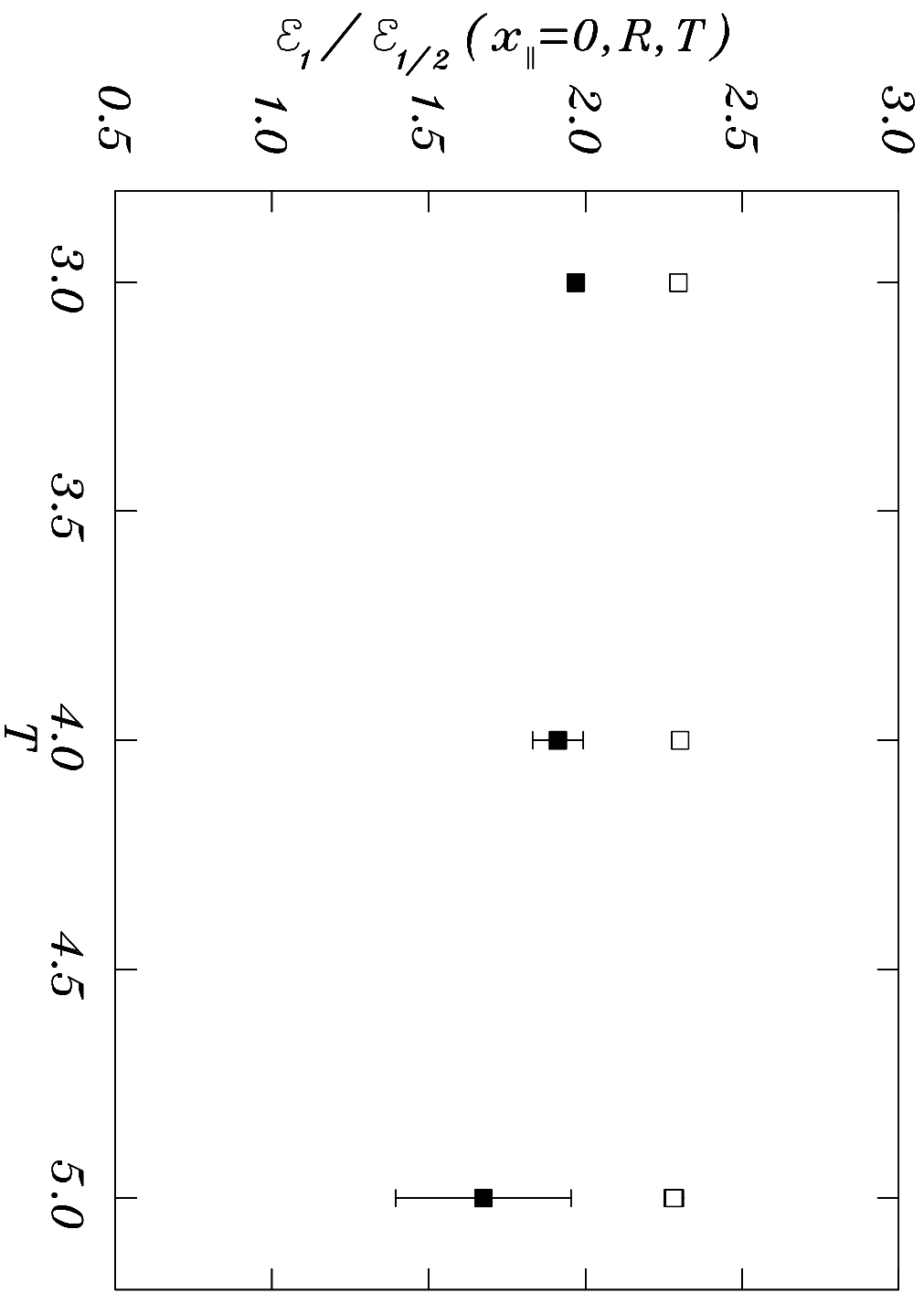

FIGURE 2

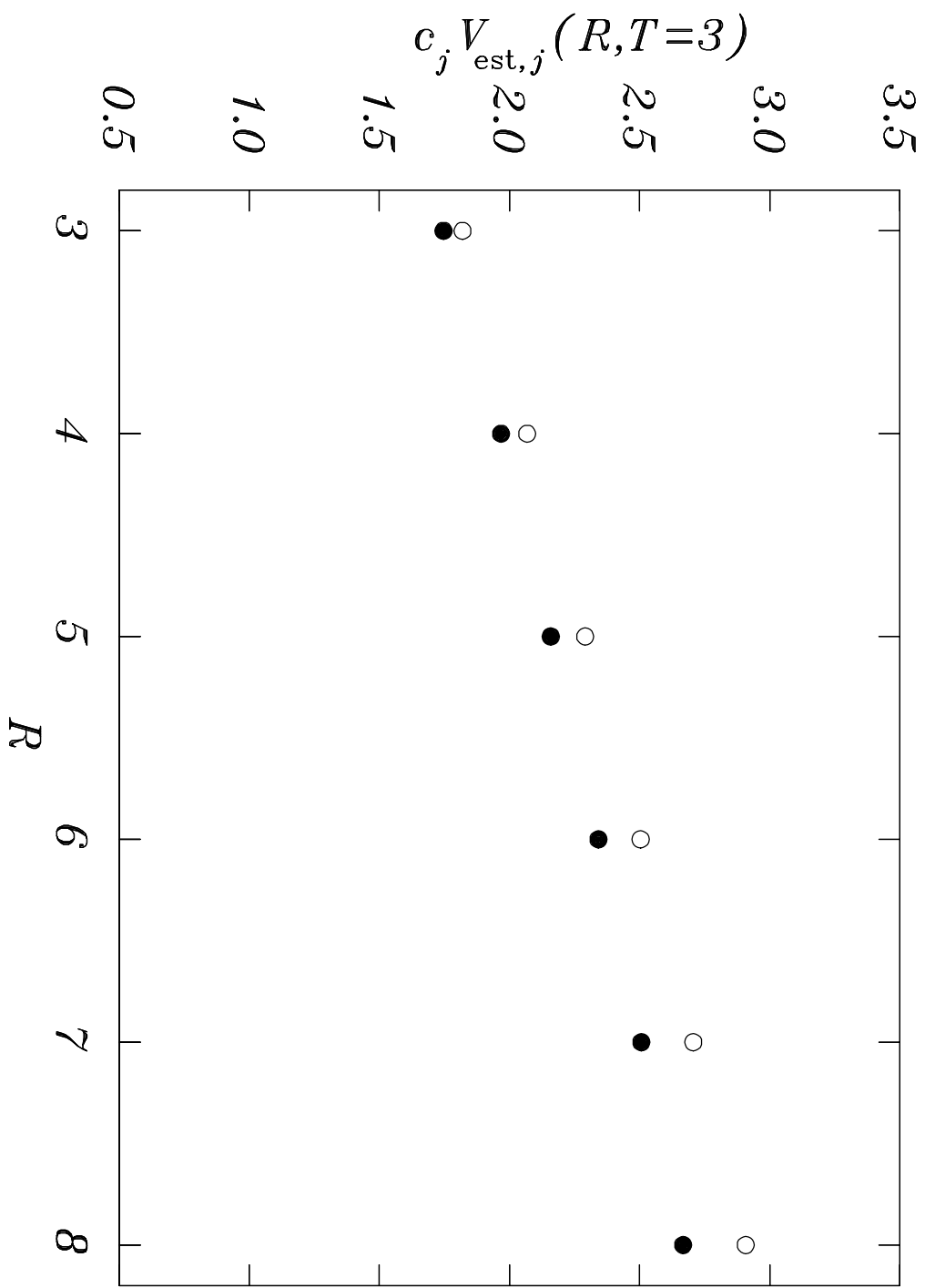

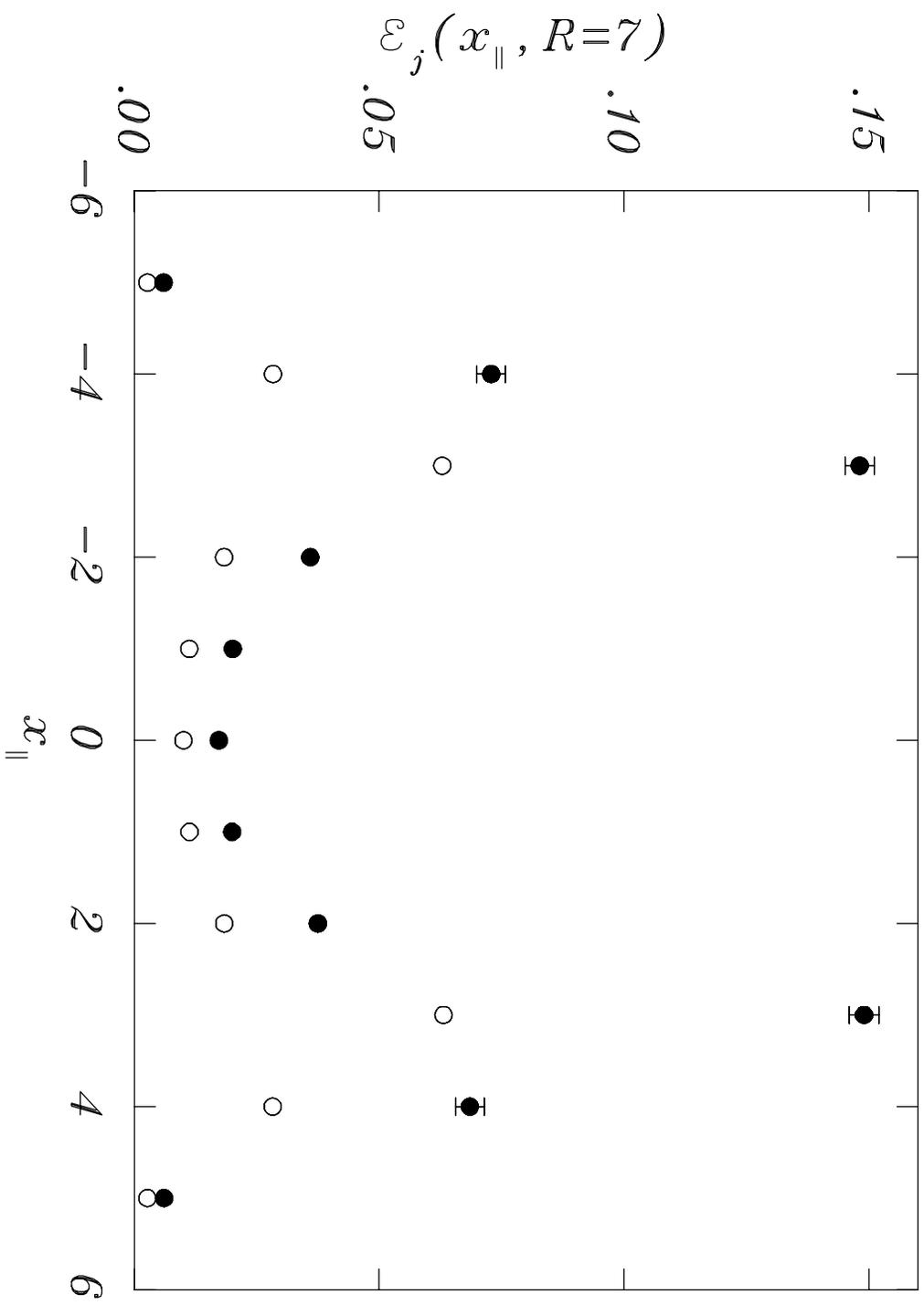

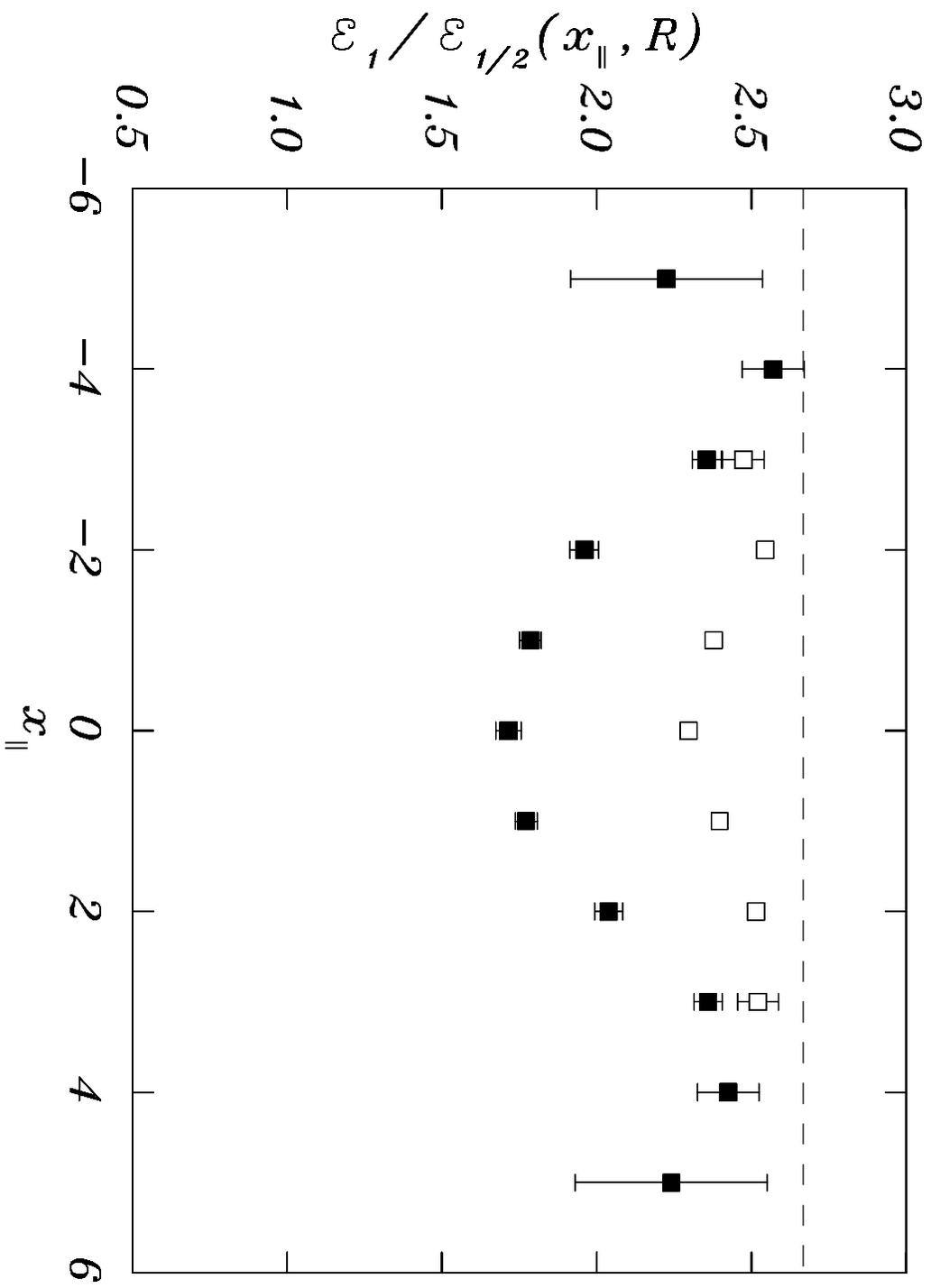

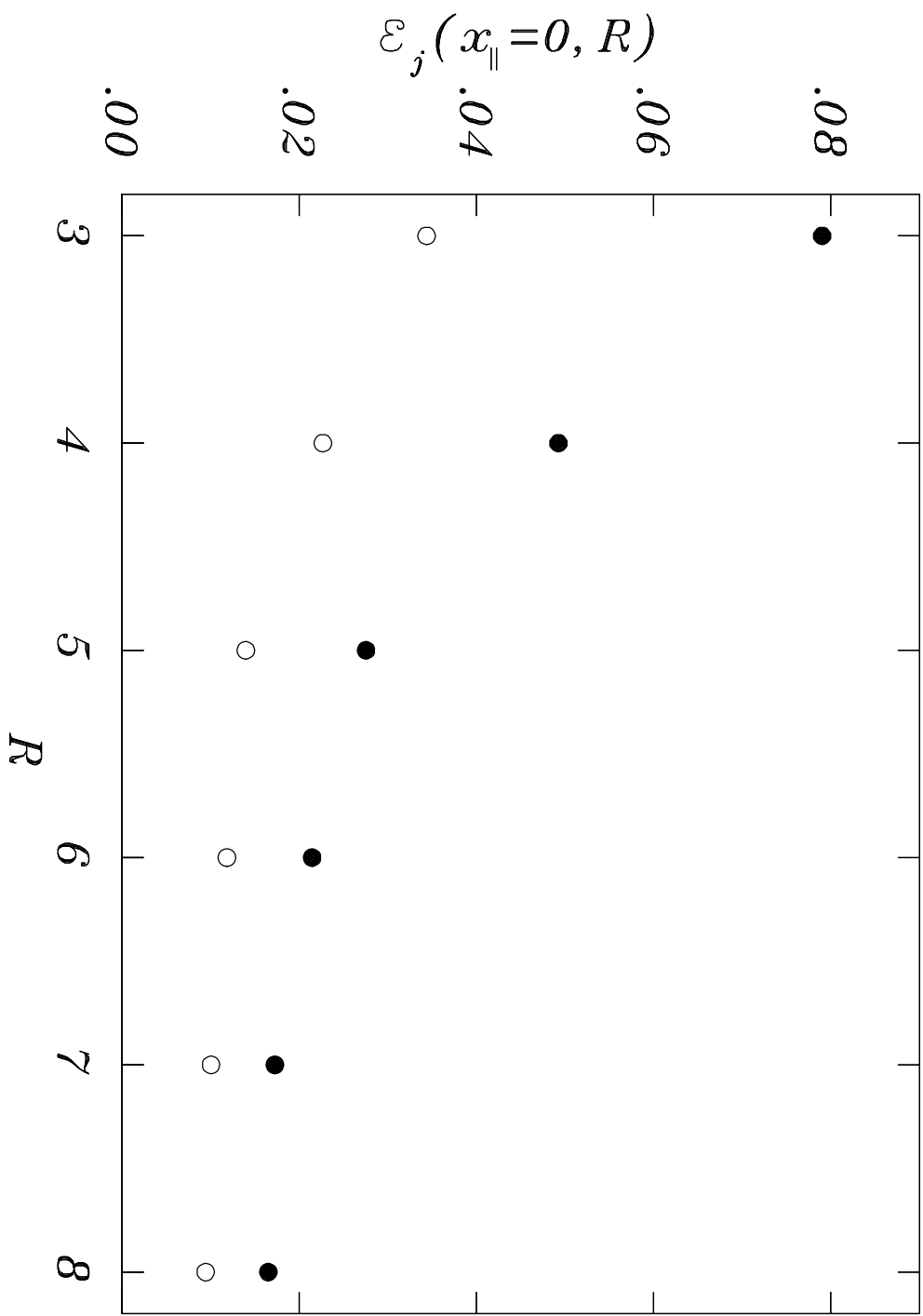

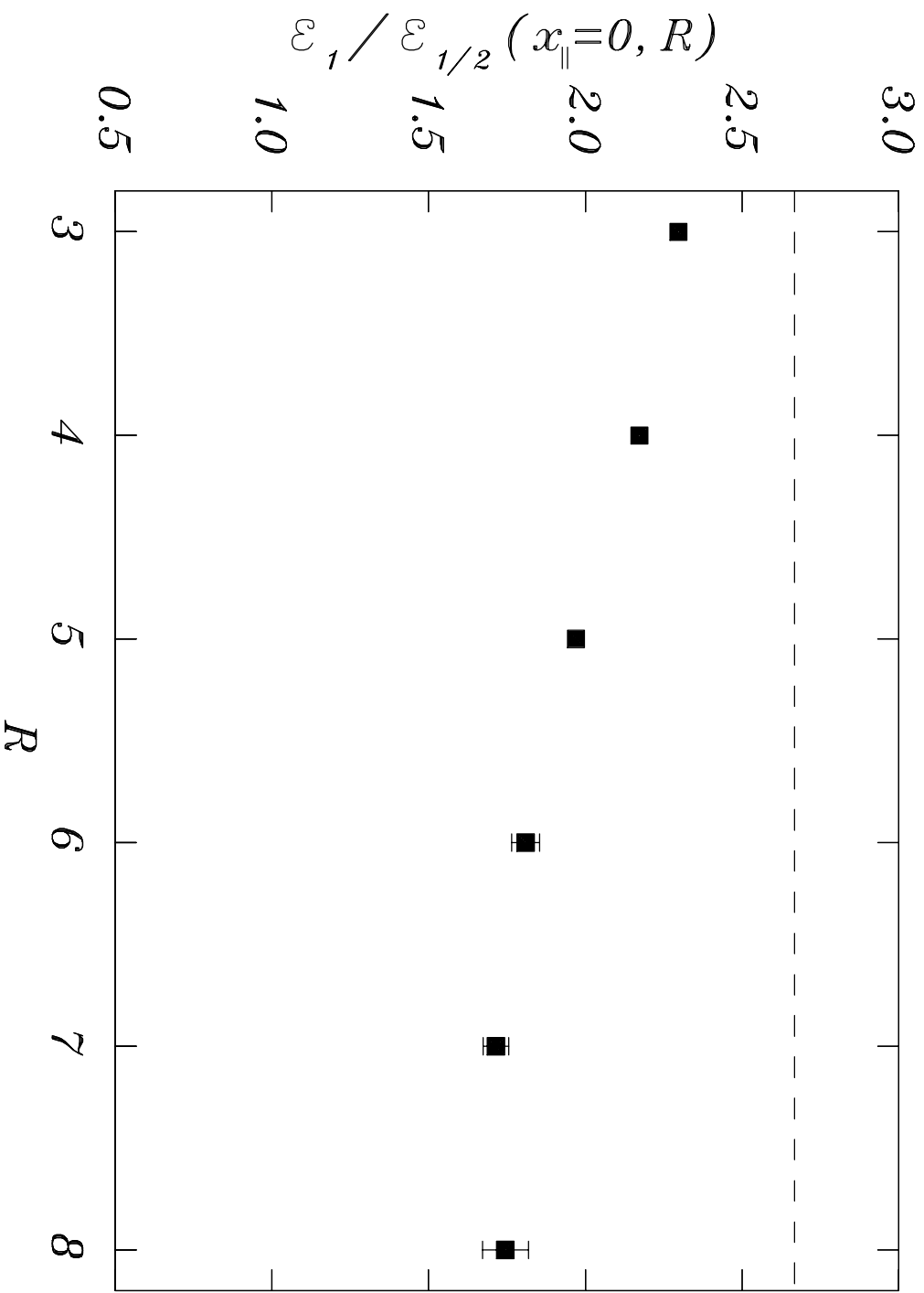

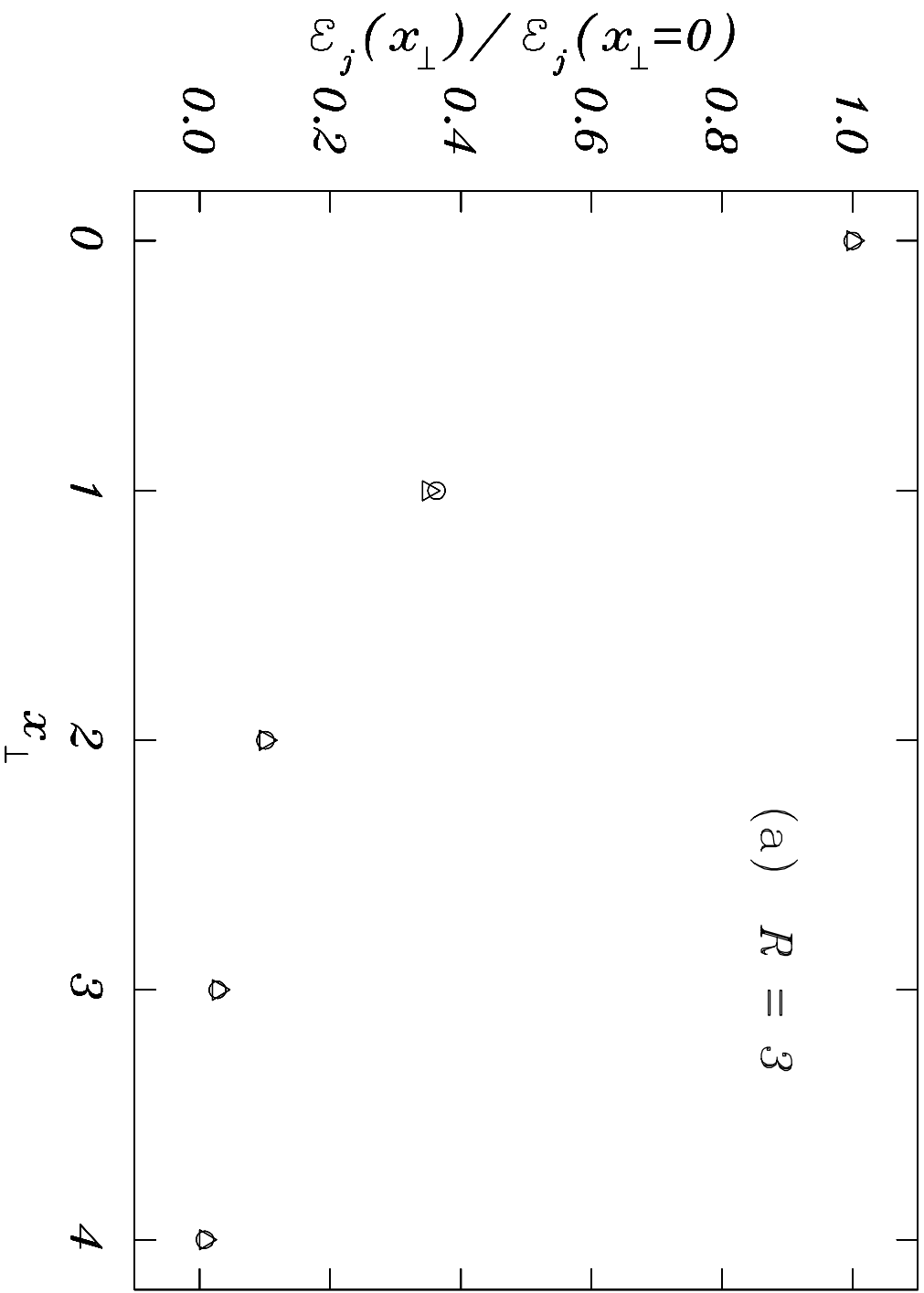

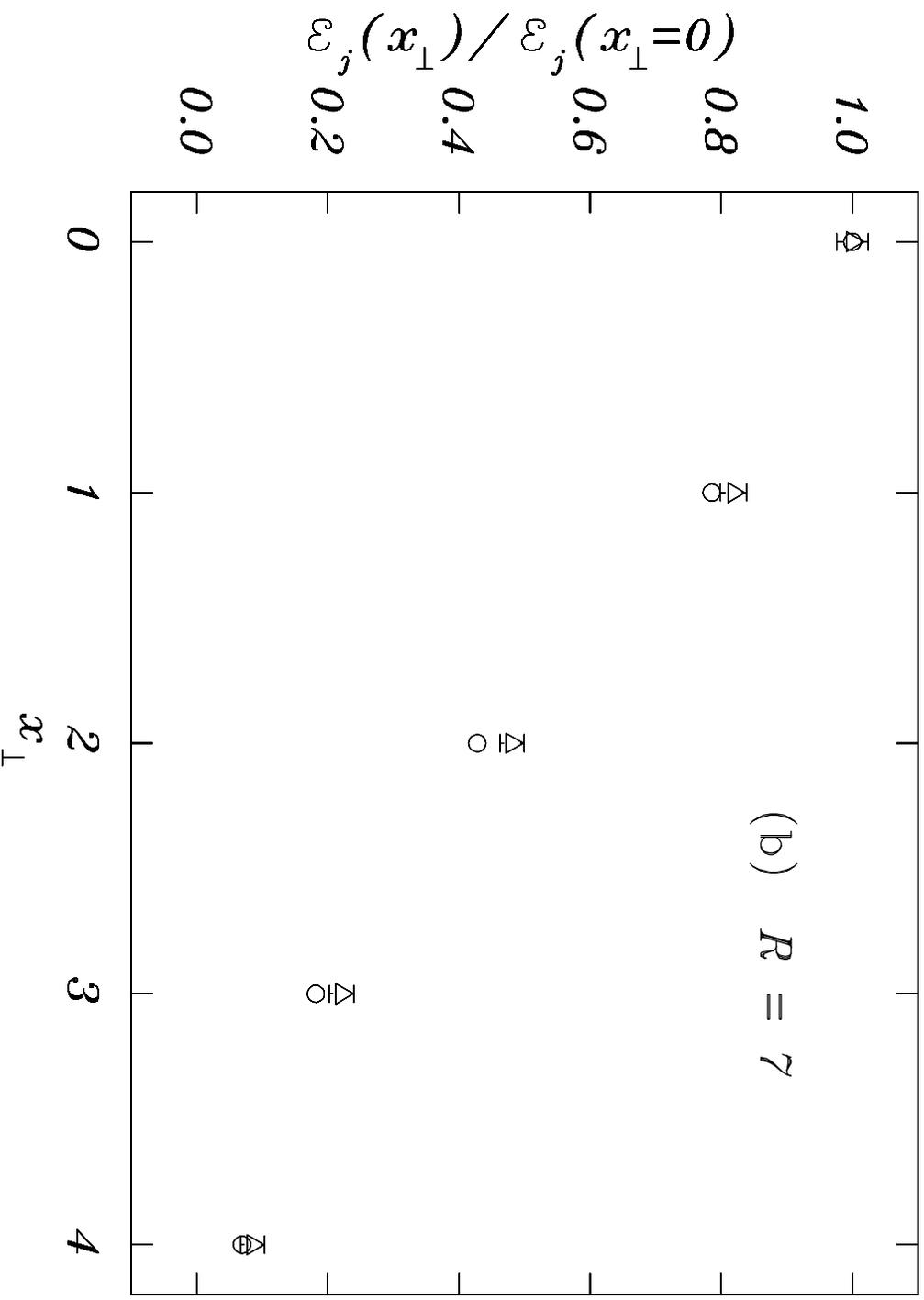

# Adjoint "quark" color fields in four-dimensional lattice gauge theory: Vacuum screening and penetration

Howard D. Trottier

*Department of Physics, Simon Fraser University, Burnaby, B.C., Canada V5A 1S6*[*]

(March 1995)

## Abstract

The fields generated by "quarks" in the adjoint representation of SU(2) color are analyzed in the scaling region of the four-dimensional lattice theory. Evidence of vacuum screening of adjoint quarks is obtained from a comparison of quark-antiquark ($Q\overline{Q}$) flux-tubes for quarks in the adjoint ("isospin" $j = 1$) and fundamental ($j = 1/2$) representations. The component $\mathcal{E}_j$ of the color-electric field strength in the direction parallel to the $Q\overline{Q}$ axis is calculated. Near the quarks the ratio of fields $\mathcal{E}_{j=1}/\mathcal{E}_{j=1/2}$ approaches the value $8/3$, which is equal to the ratio of SU(2) Casimirs. In between the quarks, the ratio falls well below $8/3$ at large $R$. $\mathcal{E}_j$ also falls off rapidly as a function of distance $x_\perp$ perpendicular to the $Q\overline{Q}$ axis. However, the ratio $\mathcal{E}_{j=1}/\mathcal{E}_{j=1/2}$ depends very weakly on $x_\perp$. The flux-tubes in the two representations thus appear to have very similar cross-sections. This result could imply that the QCD vacuum is dual to a type I superconductor.

---

[*]Email address: trottier@sfu.ca.

*Introduction* Linear confinement in QCD has been well established by lattice simulations of the quark-antiquark ($Q\overline{Q}$) potential [1]. However the physical mechanism underlying confinement has yet to be established, and this continues to represent a fundamental challenge to our understanding of QCD.

Valuable information has been obtained from studies of the color fields generated by a $Q\overline{Q}$ pair, as this provides a more detailed probe of the system than is obtained from the potential. For example, lattice simulations have demonstrated flux-tube formation [2–5]: color fields generated by quarks are found to be "squeezed" into a cylinder whose cross-section is roughly independent of their separation $R$. Flux-tube formation provides an attractive description of confinement in QCD, and is the basis of many phenomenological models [6].

This paper presents the first calculation of flux-tubes generated by adjoint "quarks" in the scaling region of four-dimensional SU(2) lattice gauge theory. It is widely expected that adjoint quarks should exhibit dynamics with some important differences from quarks in the fundamental representation, thus providing a unique probe of confinement physics [7–10,4,11]. In fact, there should be no confinement of adjoint quarks: at large separations $R$ it should be energetically favorable for the flux-tube to fission, with the formation of a pair of color-neutral quark-gluon bound states ("gluelumps") [8,9].

Recently, Michael has obtained some indication that the potential for adjoint quarks may saturate at very large $R$, where the energy in the flux-tube exceeds twice the energy of the gluelump [9] (see also Ref. [11]). Further evidence of adjoint quark screening is obtained here by comparing the distribution of fields generated by quarks $Q_j$ in both the adjoint ("isospin" $j = 1$) and fundamental ($j = 1/2$) representations of SU(2) color. A number of highly efficient techniques are exploited to maximize the quality of the results.

*Method* The SU(2) lattice theory with standard Wilson action is used. An efficient scheme for the introduction of a static $Q_j\overline{Q}_j$ pair is to use a nonlocal (or "fuzzy") Wilson loop, one that is "spread out" over an extended region that is, in principle, related to the physical dimensions of the state [12,13,8,9,1,5]. Given a set of links $U_\mu(x)$, a corresponding set of "fuzzy" links $\tilde{U}_\mu^1(x)$ is constructed according to [13]:

$$\tilde{U}_{\mu \neq 4}^1(x) = \mathcal{N}\left[c\,U_\mu(x) + \sum_{\nu \neq \pm\mu, \pm 4} U_\nu(x)U_\mu(x + \hat{\nu})U_\nu^\dagger(x + \hat{\mu})\right], \tag{1}$$

where $c$ is a positive constant, and $\mathcal{N}$ is an arbitrary normalization (conveniently chosen so that $\det \tilde{U}^1 = 1$). This procedure can be iterated. The number of iterations and the parameter $c$ are chosen "empirically" so as to maximize the overlap of a fuzzy operator with the state of interest. In order to preserve a transfer matrix fuzzing is applied only to spatial links ($\mu \neq 4$), which are mixed only with their spatial "staples" [13]. Gauge-invariant (fuzzy) Wilson loops are built in the usual way:

$$\widetilde{W}_j(R, T) \equiv \frac{1}{2j + 1}\text{Tr}\left\{\prod_{l \in L} \mathcal{D}_j[\tilde{U}_l]\right\}, \tag{2}$$

where $L$ denotes the closed loop, and $\mathcal{D}_j[\tilde{U}_l]$ is an appropriate irreducible representation of the link ("fuzzy" spatial links and "ordinary" time-like links). $\widetilde{W}_1$ can be expressed in terms of $\widetilde{W}_{1/2}$, $\widetilde{W}_1 = (4\widetilde{W}_{1/2}^2 - 1)/3$.

Lattice measurements of the $Q_j \overline{Q}_j$ field strengths are obtained from plaquette correlators. Results are reported here for the component of the color-electric field in the direction $\hat{x}$ parallel to the $Q_j \overline{Q}_j$ axis

$$\mathcal{E}_j(x) \equiv -\frac{\beta}{a^4} \left[ \frac{\left\langle \widetilde{W}_j \frac{1}{2} \text{Tr} \, U_{14}(x) \right\rangle - \left\langle \widetilde{W}_j \right\rangle \left\langle \frac{1}{2} \text{Tr} \, U_{14} \right\rangle}{\left\langle \widetilde{W}_j \right\rangle} \right], \tag{3}$$

where $U_{14}$ is a plaquette with sides along the $\hat{x}$ and time-like axes. In the continuum limit $\mathcal{E}_j$ reduces to the expectation value of $\frac{1}{2} \sum_a (\vec{E}^a \cdot \hat{x})^2$ in the presence of a $Q_j \overline{Q}_j$ pair, after vacuum subtraction [2]. In most phenomenological models the structure of the flux-tube is determined mainly by $\vec{E}^a \cdot \hat{x}$. This was shown explicitly for $j = 1/2$, 1 and 3/2 flux-tubes in three-dimensional SU(2) [4]. Lattice calculations also indicate that the potential for $j = 1/2$ quarks in four dimensions comes mainly from this field [2].

An analytical integration is made on links which occur linearly in the observable (in this case, on the time-like links in $\widetilde{W}_j$) [14]:

$$\int [dU_l] \mathcal{D}_j[U_l] e^{-\beta S} = \frac{I_{2j+1}(\beta k_l)}{I_1(\beta k_l)} \mathcal{D}_j[V_l] \int [dU_l] e^{-\beta S}, \tag{4}$$

where $S$ is the action, and $k_l V_l$ is equal to the sum of the six "staples" coupling to the link $U_l$ (det $V_l \equiv 1$ and $k_l \geq 0$). An analytical integration is also performed over the four links in a plaquette [15]:

$$\int \left[ \prod_{i=1}^{4} dU_i \right] \frac{1}{2} \text{Tr} \left( U_4^\dagger U_3^\dagger U_2 U_1 \right) e^{-\beta S} = \frac{Z'(\beta)}{Z(\beta)} \int \left[ \prod_{i=1}^{4} dU_i \right] e^{-\beta S}, \tag{5}$$

where

$$Z(\gamma) \equiv \sum_{n=1}^{\infty} n \frac{I_n(\gamma)}{\gamma} \chi_n \left( J_4^\dagger J_3^\dagger J_2 J_1 \right) \prod_{i=1}^{4} I_n(j_i \beta) \tag{6}$$

and $Z'(\gamma) \equiv dZ(\gamma)/d\gamma$. The matrix $j_i J_i$ is equal to the sum of the five "staples" connected to the link $U_i$, excluding the staple which forms part of the plaquette $U_4^\dagger U_3^\dagger U_2 U_1$ (det $J_i \equiv 1$, and $j_i \geq 0$). $\chi_n(X)$ is the character of the SU(2) matrix $X$ in the representation of dimension $n$. The characters are generated from the recursion relation $\chi_n = \chi_{n-1} \chi_2 - \chi_{n-2}$, starting from $\chi_1 \equiv 1$ and $\chi_2(X) = \frac{1}{2} \text{Tr} X$ [2]. The series in Eq. (6) converges very rapidly (at $\beta = 2.4$ less than ten terms suffice in practice). An efficient downwards recursion technique is used to generate the Bessel functions $I_n$.

Equation (5) provides for a reduction in errors in $\mathcal{E}_j$ by a factor of as much five. It is also advantageous to use $\langle W_j \rangle \langle U \rangle \simeq \langle W_j U(x_R) \rangle$ in Eq. (3) [2], where the reference point $x_R$ was taken to be 8 lattice units in each spatial direction relative to the center of the Wilson loop (which for present purposes is sufficient to ensure a good approximation, within statistical errors). This tends to reduce the effect of fluctuations in the Wilson loop [2].

*Results* The calculations were done on a $16^4$ lattice at $\beta = 2.4$. More than 10,000 sweeps were used for thermalization. 1,800 measurements were made of fuzzy Wilson loops of sizes $R \times T$ from $3 \times 1$ to $8 \times 5$. 6,000 further measurements were then made of loops with $R = 7$ and 8. 100 heat bath sweeps were made between measurements, yielding integrated autocorrelation times $\tau_{\text{int}} \lesssim 0.5$. Estimates of the statistical errors were obtained using the jackknife method. However, measurements of different observables tend to be strongly correlated if they were measured simultaneously on a given lattice.

Twenty iterations of the fuzzing procedure Eq. (1) were used, with $c = 2.5$. Figure 1 shows some results for the "time-dependent" estimates of the potential:

$$V_{\text{est},j}(R, T) \equiv -\ln\left[\frac{\widetilde{W}_j(R, T)}{\widetilde{W}_j(R, T-1)}\right]. \tag{7}$$

$V_{\text{est},j}(R, T)$ becomes independent of $T$ for $T \gtrsim 3$, within statistical errors. The same is true of $\mathcal{E}_j(R, T)$, although the statistical errors in the fields are much larger than in the potential (useful results for $R = 7$ and 8 could only be obtained for $T = 3$). Figure 2 shows the time-dependence of the fields at the center of the Wilson loop for $R = 3$ and 5.

Figure 3 shows evidence of screening in the adjoint potential that becomes more pronounced as $R$ is increased. At $R \times T = 8 \times 4$, for example, $V_{\text{est},1}/V_{\text{est},1/2} = 2.47 \pm 0.02$. Michael has obtained results for larger $R$ [9]; at $R = 12$, $V_1/V_{1/2}$ is between about 2.1 and 2.3 (the uncertainty comes from an estimate of the systematic errors in the extrapolation to $T \to \infty$). These results are to be compared with the ratio of Casimirs $j(j + 1)$ of the two representations, equal to $8/3 = 2.66$ (the Casimir can be interpreted as the squared "charge" of the quark [6]).

Screening more evident in the field strength. Figures 4 and 5 show the fields as functions of position $x_\parallel$ in the plane of the Wilson loop, using data at $T = 3$. The quarks are located at $x_\parallel = \pm R/2$. For odd $R$ the centroids of the plaquettes used in the correlator Eq. (3) lie at integer values of $x_\parallel$. Figures 6 and 7 show the fields at the center of the Wilson loop as functions of $R$ (the data for even $R$ are averaged over plaquettes with centroids at $x_\parallel = \pm 0.5$).

Near the quarks, the ratio of field strengths $\mathcal{E}_{j=1}/\mathcal{E}_{j=1/2}$ is close to the Casimir ratio $8/3$, for all $R$. In the region between the two quarks the ratio of fields falls below $8/3$, reaching a minimum at $x_\parallel = 0$. The ratio is significantly reduced at large $R$. At $R = 5$, for example, $\mathcal{E}_{j=1}/\mathcal{E}_{j=1/2} = 1.97 \pm 0.02$ at $T = 3$, and $1.91 \pm 0.08$ at $T = 4$. Quantitatively similar results for the ratio of fields at a given $R$ are found at all points in the plane perpendicular to the $Q_j\overline{Q}_j$ axis (see Figs. 8(a) and (b), described below). This suggests that there is a reduction in the color-electric flux in the region between the adjoint quarks.

Some care is required in comparing the plaquette correlator data at fixed $T$ for different $R$, since the systematic errors in the $T \to \infty$ extrapolation are expected to increase rapidly with $R$ (the use of fuzzed Wilson loops to excite the $Q_j\overline{Q}_j$ pair becomes less efficient as $R$ increases). The statistical errors for large $R$ and $T$ are too big to permit a reliable estimate of systematic errors in the extrapolation. Nevertheless, it appears that in some spatial regions the local adjoint field strength exhibits more pronounced screening than is indicated by the potential. This is presumably due to the fact that the potential averages the fields over all space.

One naively expects that adjoint quarks become completely screened at sufficiently large $R$, where it becomes energetically favorable for the flux-tube to fission, leading to the formation of a pair of quark-gluon bound states ("gluelumps") [8,9]. One would therefore expect the fields in between the quarks to approach vacuum values at large $R$. A calculation on larger lattices of the plaquette correlators at large $R$ and $T$ could shed light on this question. Unfortunately, the methods used here appear to be inadequate for this task.

Results for $\mathcal{E}_j$ as a function of position $x_\perp$ perpendicular to the $Q_j\overline{Q}_j$ axis are shown in Figs. 8(a) and (b) (data was averaged over equidistant points along the two lines $\hat{y}$ and $\hat{z}$ from the center of the loop). The fields exhibit a "penetration depth" of a few lattice spacings. The fundamental and adjoint quark flux-tubes appear to have very similar cross-sections, despite appreciable screening of the fields in between the adjoint quarks at large $R$.

This result has implications for the dual superconductor model of confinement [16], which is motivated by an analogy to compact Quantum Electrodynamics (QED), where confinement arises from vacuum condensation of magnetic monopoles [17]. In type II superconductors a domain wall formed between normal and superconducting regions has negative surface energy, so that many normal regions are created, each carrying an elementary quantum of flux. In type I superconductors on the other hand a domain wall has positive surface energy, hence as few normal regions as possible are created [18]. If the dual superconductor picture is correct, then the fact that adjoint quarks form a single flux-tube, with a cross-sectional structure similar to the flux-tube formed by fundamental quarks, may suggest that the QCD vacuum is dual to a type I superconductor. Evidence for dual supercurrents associated with domain wall formation in SU(2) lattice gauge theory has recently been reported in Ref. [19], but the type of superconductivity could not be distinguished.

These results also suggest a connection between the confinement mechanism in QCD in both three and four dimensions, and in three-dimensional QED. The fields generated by $Q\overline{Q}$ pairs in several representations of three-dimensional SU(2) and U(1) lattice theories were calculated in Ref. [4]. The fields were found to be restricted to a flux-tube whose cross-section is approximately independent of the representation. It is interesting to note that Abelian magnetic monopoles are present in all of these theories.

## ACKNOWLEDGMENTS

The author is grateful to Richard Woloshyn and Greg Poulis for fruitful conversations. This work was supported in part by the Natural Sciences and Engineering Research Council of Canada.

## FIGURES

FIG. 1. Time-dependence of the estimated potential at $R = 7$ for $j = 1$ ($\bullet$) and $j = 1/2$ ($\circ$). Statistical errors are shown when bigger than the plotted symbols.

FIG. 2. Time-dependence of the ratio of fields $\mathcal{E}_{j=1}/\mathcal{E}_{j=1/2}$ at the center of Wilson loops with $R = 3$ ($\square$) and $R = 5$ ($\blacksquare$).

FIG. 3. Estimated potential versus quark separation $R$ for $j = 1$ ($\bullet$) and $j = 1/2$ ($\circ$). The $j = 1/2$ potential has been rescaled by $c_{1/2} \equiv 8/3$ $(c_1 \equiv 1)$.

FIG. 4. Field strength versus position $x_\parallel$ in the plane of the Wilson loop with $R \times T = 7 \times 3$, for $j = 1$ ($\bullet$) and $j = 1/2$ ($\circ$).

FIG. 5. Ratio of fields $\mathcal{E}_{j=1}/\mathcal{E}_{j=1/2}$ versus $x_\parallel$ for Wilson loops with $R = 3$ ($\square$) and $R = 7$ ($\blacksquare$). In both cases $T = 3$. The dashed line shows the ratio of Casimirs.

FIG. 6. Fields $\mathcal{E}_j$ in the center of the Wilson loop versus $R$ for $j = 1$ ($\bullet$) and $j = 1/2$ ($\circ$). The data are at $T = 3$.

FIG. 7. Ratio of fields $\mathcal{E}_{j=1}/\mathcal{E}_{j=1/2}$ in the center of the Wilson loop versus $R$.

FIG. 8. Fields $\mathcal{E}_j$ versus position $x_\perp$ perpendicular to the $Q_j \overline{Q}_j$ axis for $j = 1$ ($\triangle$) and $j = 1/2$ ($\circ$). The fields have been normalized to their values at $x_\perp = 0$. The results are shown for (a) $R = 3$ and (b) $R = 7$. The data are at $T = 3$.